# Title

Calculating Great Britain's half-hourly electrical demand from publicly available data.

# Authors

IA Grant Wilson[1], Shivangi Sharma[1], Joseph Day[1], Noah Godfrey[1]

**Affiliations**

[1] Energy Informatics Group, The University of Birmingham, UK

corresponding author(s): Dr Grant Wilson (i.a.g.wilson@bham.ac.uk)

IA Grant Wilson: Conceptualization, Methodology, Software, Data curation, Writing- Original draft preparation, Writing- Reviewing and Editing.:  Shivangi Sharma: Writing- Reviewing and Editing.: Joseph Day: Writing- Reviewing and Editing. Noah Godfrey: Software, Data curation, Writing- Reviewing and Editing.

# Abstract

Here we present a method to combine half-hourly publicly available electrical generation and interconnector operational data for Great Britain to create a timeseries that approximates its electrical demand. We term the calculated electrical demand 'ESPENI' that is an acronym for <u>E</u>lexon <u>S</u>um <u>P</u>lus <u>E</u>mbedded <u>N</u>et <u>I</u>mports. The method adds value to the original data by combining both transmission and distribution generation data into a single dataset and adding ISO 8601 compatible datetimes to increase interoperability with other timeseries data. Data cleansing is undertaken by visually flagging data errors and then using simple linear interpolation to impute values to replace the flagged data. Publishing the method allows it to be further enhanced or adapted and to be considered and critiqued by a wider community. In addition, the published raw and cleaned data is a valuable resource that saves researchers considerable time in repeating the steps presented in the method to prepare the data for further analysis. The data is a public record of the decarbonisation of Great Britain's electrical system since late 2008, widely seen as an example of rapid decarbonisation of an electrical system away from fossil fuel generation to lower carbon sources.

# Introduction

Great Britain's electrical generation mix has seen significant change from 2009 through to 2021, with an increase in wind, solar and biomass generation and a decrease from fossil-fuels. Most important from the perspective of greenhouse gas emissions reduction has been the near complete shift away from coal generation over the 2015-2019 5-year period. A major contributory factor facilitating this change was the reduction in the overall electrical demand by about 67 TWh (20%) between 2010 and 2020. Without this reduction, the extra electrical demand would have been satisfied by marginal generation plant available to the system or imports and allowed coal generation greater opportunities to remain on the system. Having the data to evidence the change in Great Britain's electrical demand and generation mix is important from both a research and an educational perspective. This article describes a method to combine two publicly available datasets to provide one avenue that provides that evidence.

Figure 1 shows the annual electrical energy in TWh from different generation types from 2009. The data that underlies Figure 1 are publicly available from two separate data sets with a time granularity of 30 minutes, but neither dataset contains an ISO 8601[1] compatible date and time format, e.g., 2020-07-11T13:00:00Z . The datasets thus require effort to parse into a more interoperable and useable format for further analysis and visualisation and a method to do so is presented in this article. There is evidence that adding value to the raw data using this method is useful, as the cleaned data with the ISO 8601 compatible datetimes have been downloaded from the Zenodo platform[2] over 1400 times in the initial twelve months of being openly available.

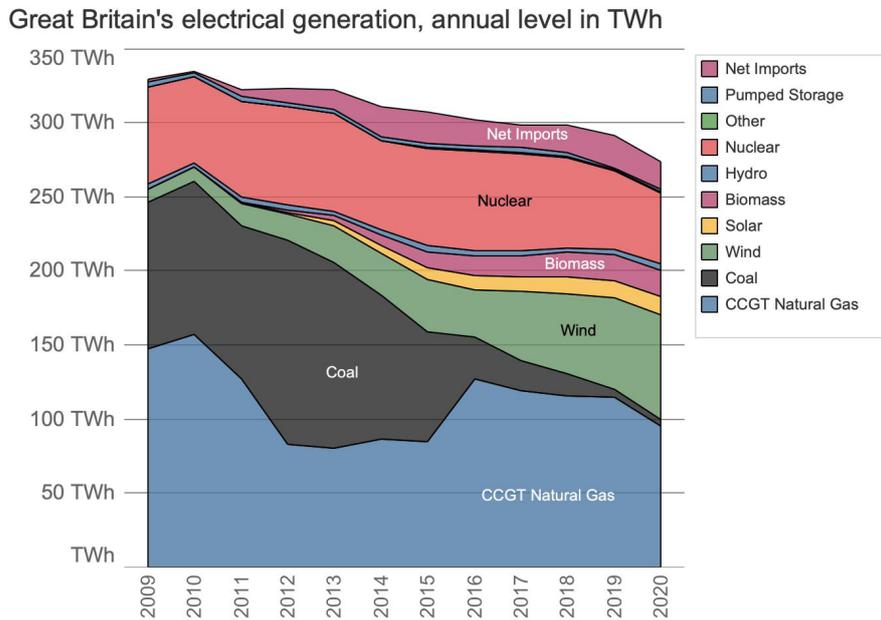

*Figure 1 - Great Britain's changing fuel supplies for electrical generation from 2009 – 2020: aggregated to annual values*

In all national electrical systems data from generation and demand are required to allow the system operator to monitor and balance the electrical network in real-time. In Great Britain, larger electrical generation plants and interconnectors are connected directly to the higher voltage electrical transmission system and they have to agree to a code of operation defined in the balancing and settlement code[3,4]. In addition, companies that provide significant electrical demand on the system, such as suppliers to end users, also must sign up to the balancing and settlement code. A major element of the code is to mandate that all parties are required to inform the system operator (National Grid) of their planned generation or demand in **advance** of real-time, so it can take various measures to keep the electrical system in balance. The data presented here is the aggregation of the operational data over a 30-minute Settlement Period. The operational data is far more granular and is used by the system operator to keep the electrical system in balance, however, it is not publicly available and therefore not the subject of this article.

In Britain, the electrical market has distinct blocks where balancing is managed on a rolling basis, and this is termed a Settlement Period (the 'period' over which the balancing actions are 'settled'). There are 48 half-hourly Settlement Periods in a standard day. Due to clock changes for British Summer Time, in the last Sunday in March there are two less Settlement Periods (46) as the clocks are changed forward an hour, and therefore an hour is 'lost' from the settlement day in localtime. In the last Sunday of October there are two more Settlement Periods (50) as clocks are changed back an hour and therefore an extra hour is 'gained' by the settlement day in localtime.

Informing the system operator of the estimated generation or demand positions of all parties at least an hour in advance of real-time allows it to coordinate balancing activities to happen prior to the start of the real-time half-hourly Settlement Period itself. This takes place on a rolling basis, so that each distinct Settlement Period is managed separately.

During the Settlement Period, the electrical system operator has various additional mechanisms and contracts that it uses to increase or decrease supply or demand on the electrical system to keep it in balance at all times. If there is too much generation and not enough demand the system frequency will increase, and it will drop if generation is lower than demand. The system operator has a mandatory duty as set out in clause CC.6.1.2 in the Grid Code[5] stated as: *'The Frequency of the National Electricity Transmission System shall be nominally 50Hz and shall be controlled within the limits of 49.5 -50.5Hz unless exceptional circumstances prevail'*.

Contained within the advance information given to National Grid of the balancing mechanism parties' forecast generation and demand, is additional data such as the price and amount of deviation that the balancing mechanism party would be prepared to shift from its forecast generation or demand. In essence, the balancing mechanism party can submit not only its forecast position, but also how it is prepared to increase or decrease its forecast supply or demand from that forecast position. This means that National Grid can compare and accept various prices and actions to prepare the Settlement Period in advance, to better match the levels of forecast generation with the value that National Grid has estimated the demand is likely to be. Any adjustments to forecast initial positions of generation or demand are therefore known and agreed in advance of the real-time of the Settlement Period between National Grid and the balancing mechanism party.

The adjustments agreed prior to and during a Settlement Period will have a cost. A principle that is applied is that any party that has caused a system imbalance will pay an amount proportionate to the scale of the imbalance that they have caused. This will be a fraction of the total cost of system balancing for that particular Settlement Period. In essence, this is a financial penalty levied on those parties that cause an imbalance in comparison to their peer group that do not. In a competitive environment such as Britain's regulated electrical system there is a financial self-interest to more accurately forecast generation or demand or reduce the amount of imbalance that a party might regularly cause. This in theory should minimise the balancing effort that National Grid would require to take, thus reducing the overall cost of system balancing.

To account for the financial allocation of the costs of a Settlement Period, the relevant data are sent to a wholly owned subsidiary of National Grid called Elexon. They then use these data to bill the parties that caused an imbalance and pay those parties that have helped to reduce an imbalance. In simple terms, National Grid's focus as the Electrical System Operator is on keeping the generation and demand of electrical energy balanced. Then, post-hoc, Elexon manages the flow of funds between generation or demand parties that caused an imbalance and those that helped to bring the system back into balance. Elexon make some of this half-hourly data publicly available through their data platform[6] that can be accessed after registering with the platform. Elexon's aggregated generation by fuel type is the first of the datasets used in the method presented in this article to approximate Great Britain's electrical demand at a half-hourly level. The data presented here is therefore an outcome of the Balancing Mechanism where the aggregation and reporting is defined by the window of a Settlement Period of 30-minutes. All electrical grids have to manage operational balancing on a sub-second basis by a system operator, but it aggregating to a time-window to facilitate reconciliation of costs is self-evidently useful approach. Different electrical markets can have different time-windows for control and reconciliation, e.g., Germany and Switzerland have a 15-minute market window[7].

The method presented here uses Elexon data that details electrical generation connected at the transmission level, which are monitored as part of the electrical system's balancing mechanism. The transmission level is the higher voltage level utilised to transmit bulk levels of electricity over longer distances; larger generators or industrial users connect directly to the transmission level. The method then combines these Elexon data with estimates for embedded/distributed solar and wind generation from the system operator National Grid. Electrical distribution grids connect the transmission level electrical grid to end users and are at a lower voltage. Smaller generators and users will be connected to the distribution grid, rather than directly at a higher voltage to the transmission grid. The resulting timeseries data therefore has both transmission connected and embedded/distribution connected generation, which overcomes one of the limitations of solely using the transmission connected Elexon data to provide a more accurate representation of Great Britain's electrical demand. The generation values are presented in their original as well as a visual error checked and imputed dataset; both have been parsed to provide coordinated universal time (UTC) and localtime values in ISO 8601 compatible format which overcomes a further limitation of the original data that has settlement period values as a proxy for datetime values. The method has been developed and adapted over several years as the categories of the underlying raw data have themselves been amended by the data providers.

After login, the data that Elexon makes publicly available can be downloaded either manually[8], or using an application programming interface with an alphanumeric code assigned to the registration account using the Balancing Mechanism Reporting Service[9]. As an original source of publicly available electrical system data at a half-hourly level Elexon's data are heavily utilised to underpin numerous detailed research and analyses of Great Britain's energy systems[10–88]; public facing websites[89–92]; and are also particularly useful in teaching and learning for energy systems courses. Aggregated fuel type data are also available at 5-minute granularity, which can also be downloaded from Elexon either manually or through an API. More granular time data brings additional advantages dependent on a research question, but it also carries a disadvantage in terms the ease of cleaning when undertaken manually. Therefore, half-hourly data was chosen to present the method as this has been the preferred level of detail within the research group at the University of Birmingham for research and for teaching and learning.

Regardless of the granularity of the Elexon data (whether 30 minute or 5-minute data), there are four major limitations of the Elexon data:

1) They do not include distributed generation for solarPV or embedded wind
2) They do not include datetime values in an ISO 8601 compatible format as the data are only provided with balancing mechanism Settlement Date and Settlement Period values
3) There are no negative values for pumped storage (charging) or interconnectors (exporting)
4) They only include generation from parties that are part of the balancing mechanism and thus do not include various autogeneration such as many combined heat and power plants.

The method presented here addresses the first three limitations by combining data from another publicly available dataset from National Grid through their 'historic demand data' webpage[93], and by adding ISO 8601 compatible datetime values. The additional dataset from National Grid contains the categories 'EMBEDDED_SOLAR_GENERATION' and 'EMBEDDED_WIND_GENERATION' as well as having negative values for interconnectors. Aggregating the estimated embedded solar and wind data with the Elexon transmission connected generation data and replacing the Elexon interconnector data with the National Grid interconnector data (which include negative values) provide a more complete picture of the generation and interconnector flows for any given half-hourly Settlement Period. An alternate solar dataset can be downloaded using an application programming interface from Sheffield Solar (https://www.solar.sheffield.ac.uk/pvlive/). However, values for distributed wind seem only to be available from National Grid.

In addition to combining the transmission connected generation from Elexon and the embedded generation from National Grid, having negative interconnector values allows exports to be considered and thus subtracted from the interconnector imports. This provides a more accurate representation of the electrical *demand* of Great Britain rather than simply its electrical generation. At times of net-export (when exports are greater than imports) the generation in Britain will be greater than its demand. In contrast, negative values for pumped storage are not subtracted from the generation total, as pumped storage demand is contained within the national boundaries of Britain.

The resulting total of Elexon generation plus the embedded/distributed solar and wind generation plus the value of net-imports approximates Great Britain's half-hourly electrical demand. This GB demand approximation is termed **ESPENI,** which is an acronym for **E**lexon **S**um **P**lus **E**mbedded **N**et **I**mports.

Presenting this methodology provides transparency and allows a comparison with more accurate methodologies as they are developed in future. It might also help a wider audience to shift to using more granular national electrical demand data. Also, regardless of the pathway ultimately taken by the Great Britain to transition to net-zero by 2050, the authors feel strongly that the analysis of future pathways benefits from a more accurate representation of its historical electrical demand. Forecasting is influenced by the initial conditions that seed the starting point of the forecasts; a more accurate representation of the initial starting point is therefore an important determinant in reducing future uncertainty.

We feel this is true for all energy systems and therefore the importance of empirical data from the operation of existing energy systems and the methodologies to curate the data are expected to continue to grow in importance. In addition, these data may help to evidence the ongoing discussion about the future of those energy systems from a wider group of stakeholders.

By publishing this methodology and data it opens up both to greater scrutiny by a wider group of research communities to consider how to improve it, and importantly to be aware of its limitations when using publicly available data. From an international perspective, we hope the basic principles contained in the method of combining transmission and distribution connected generation and netting exports from imports are useful to other electrical markets to provide a more accurate and transparent estimate of their electrical demand.

We also hope that by sharing the datasets created using the methodology, this will allow researchers to concentrate their valuable time on other aspects of data analysis, rather than having to repeat the effort in parsing data to provide an ISO compatible UTC and localtime value for each data point, combine the different years and datasets into a single csv, and replace errors with interpolated data values.

Future work will consider extending the 30-minute granularity down to 5-minute granularity and undertaking analysis using the 30-minute data on the changes to residual load and on the link between electrical demand, supply and weather patterns.

## Method

The method uses publicly available data from Elexon[8] and National Grid[94] to create the ESPENI dataset that can be downloaded from the Zenodo platform[2] under a CC-BY-NC licence.

**Elexon data**

Elexon data can be manually downloaded from the 'Generation by Fuel Type - Historic HH' webpage[8] on Elexon's portal[6]. The data can also be scripted to be downloaded from Elexon with details to be found on Elexon's website[95]. The manually downloaded Elexon data are packaged in annual files that can be added to a basket and then downloaded in bulk.

The files are comma separated text files with a header row that are encoded using the 8-bit Unicode transformation format (Utf-8).

- The oldest file is the fuelhh_2008.csv file. This contains 15 columns with the header row containing the text: #Settlement Date, Settlement Period, CCGT, OIL, COAL, NUCLEAR, WIND, PS, NPSHYD, OCGT, OTHER, INTFR, INTIRL, INTNED, INTEW
- The newest file is the fuelhh_2021.csv file, which is updated on a daily basis. This contains 20 columns with the header row containing the text: #Settlement Date, Settlement Period, CCGT, OIL, COAL, NUCLEAR, WIND, PS, NPSHYD, OCGT, OTHER, INTFR, INTIRL, INTNED, INTEW, INTELEC, INTIFA2, INTNSL, BIOMASS, INTNEM
- Additional columns from the oldest file: 'BIOMASS' appeared in the fuelhh-2017.csv file, 'INTNEM' appeared in the fuelhh_2018.csv file, INTELEC, INTIFA2 and INTNSL appeared in the fuelhh_2020.csv file.

The values in the fuel type columns are in power (MW) which is an average value over the half-hourly Settlement Period. The data start from 2008-11-05 Settlement Period 43. The data are of high quality, as there are no repeat rows and very few errors in the fuel type columns. However, as shown in Table 1 the Minimum value of all the generation values is 0, which suggests errors for categories such as CCGT and Nuclear that would not have had a zero output at a national level between 2009 and 2021.

After the individual annual files have been downloaded and saved into a folder, they can be parsed to create a single timeseries.

The annual Elexon fuelhh csv files are imported into Python using the pandas.read_csv() command in Python from the Pandas package and then combined to create a single dataframe with a row for each date and Settlement Period. There is no time data in the raw data, only the Settlement Date, and the Settlement Period. The data were then parsed to create UTC and localtime values, which allows subsequent timeseries analysis to be undertaken.

*Table 1 – Column names and descriptions for the Elexon fuelhh manual downloaded files and min and max values till 2021-05-04*

| Column header | Format | Description | Min Value | Max Value |
|---|---|---|---|---|
| #Settlement Date | YYYY-mm-dd | Date column in year-month-day format | 2009-11-05 | 2021-05-04 |
| Settlement Period | Integer | Settlement Period from 1 to 48 (note values 1-9 are single character values) | 1 | 50 |
| All values for generation listed below are aggregated by fuel type or interconnector and are the average power over the Settlement Period in MW | | | | |
| CCGT | Integer | Combined Cycle Gas Turbine generation | 0 | 27131 |
| OIL | Integer | Oil generation | 0 | 2646 |
| COAL | Integer | Coal generation | 0 | 26044 |
| NUCLEAR | Integer | Nuclear generation | 0 | 9342 |
| WIND | Integer | Transmission connected wind generation | 0 | 14095 |
| PS | Integer | Pumped storage generation. When pumped storage is a net demand (i.e., this should be a negative number) it is reported as a zero value. | 0 | 2660 |
| NPSHYD | Integer | Non pumped storage hydro generation | 0 | 1403 |
| OCGT | Integer | Open Cycle Gas Turbines | 0 | 874 |
| OTHER | Integer | Originally contained the values for BIOMASS, after 2017-11-01T20:30:00+00:00 the values no longer contained the BIOMASS category. Contains landfill generation | 0 | 2441 |
| INTFR | Integer | Interconnector France (IFA) | 0 | 3322 |
| INTIRL | Integer | Interconnector Ireland | 0 | 402 |
| INTNED | Integer | Interconnector Netherlands (BritNed) | 0 | 1132 |
| INTEW | Integer | Interconnector East West | 0 | 506 |
| INTELEC | Integer | Planned Interconnector (ElecLink) | NaN | NaN |
| INTIFA2 | Integer | Interconnector France (IFA2) | 0 | 994 |
| INTNSL | Integer | Planned Interconnector (North Sea Link) | 0 | 4 |
| BIOMASS | Integer | Biomass generation. Values started on 2017-11-01T20:00:00+00:00 prior to this they were contained within the OTHER category | 0 | 3204 |
| INTNEM | Integer | Interconnector NEMO Link | 0 | 1020 |

The initial steps to parse the data to add datetime values were:

1. Change all column header text to uppercase, to strip whitespace, to replace the # character before the Settlement date text and replace any spaces with an underscore character.
2. Change the numerical values in the SETTLEMENT_PERIOD column to text and then prepend the text character '0' to the single digit text values. This creates text values from 01 to 50 for the Settlement Periods.
3. Create an additional column called SDSP_RAW that concatenates the new values in the SETTLEMENT_DATE column and the SETTLEMENT PERIOD column using an underscore character. This creates a column of text values such as 2020-01-19_09 that is the value for the 9th Settlement Period on the 19th of January 2020. This SDSP_RAW column is useful as a text column that can then be used to sort the other columns and check for duplicate rows.
4. There should only be one row of values for each SETTLEMENT_DATE and SETTLEMENT_PERIOD and with the data from 2009-11-05 to 2021-05-04 there were no duplicate rows, which indicates the high quality of the raw data.
5. A final step was to reset the index.

Although this type of text and datetime parsing can be coded in many programming languages (e.g., R, Fortran90), the research group predominantly uses Python[96] and Pandas[97] to undertake this type of parsing due to the ease of datetime analysis and string manipulation.

Columns that are not power columns were given a prefix of 'ELEXM_' to denote that the column contained values originally from Elexon. All power columns were given a prefix of 'POWER_ELEXM_' and a suffix of '_MW' to denote that the column contained values for power, were originally from Elexon and have units of MW.

As each standard day should have 48 half-hourly Settlement Periods, a short day at the start of British Summer Time should have 46, and a long day at the end of British Summer Time should have 50; each non-leap year should have 17520 rows, with 17568 rows for a leap year. However, years 2018, 2019 and 2020 were found to have only 17519 rows, which transpired to be a missing row at the year end. These missing rows were re-created in the datetime parsing and a value was interpolated between the preceding and following values and rounded to an integer.

**Elexon data: Creating datetime columns**

The next step was to create UTC and localtime columns in ISO 8601 compatible format from the Settlement Date and Settlement Period values. A masterdatetime_iso8601.csv text file was created using a separate piece of code (https://github.com/iagw/masterdatetime.git). This csv was then used to create key:value pair dictionaries using the 'SDSP_RAW' text for every half-hourly period as the key (e.g., 2020-01-19_09) and then the ISO 8601 compatible text as the value. One dictionary had the SDSP_RAW keys matched to UTC values; another dictionary had the SDSP_RAW keys matched to localtime values. New columns were then created in the Elexon dataframe by simply mapping the SDSP_RAW key: value pairs in the dictionaries to give the UTC or localtime value. This provided a repeatable, direct one-to-one relationship that provided a robust method of creating datetime values. The values for UTC and localtime were kept as strings, and therefore any manipulation, sorting and mapping could be done using text manipulation. As the format is ISO 8601 compatible, it is simple using Python and Pandas pd.to_datetime() to change the datetime strings into time zone aware datetimes for further analysis when required.

**Elexon data: Negative values for interconnectors**

A limitation of the data from the manually downloaded Elexon source is that the values have been pre-parsed by Elexon to have a minimum floor value of zero, meaning there are no negative values in the data. This makes sense from the perspective of looking exclusively at generation, where a value should only ever be positive, but it also means that the values for exports through the electrical interconnectors cannot be calculated. The data for each of the interconnectors are therefore taken from the National Grid dataset that does contain these negative values. The interconnector columns from the Elexon data were dropped and therefore do not appear in the merged ESPENI dataset.

**Elexon data: OTHER category split into OTHER and BIOMASS categories**

On 2017-11-01 the Elexon fuel type category OTHER was split into two categories: OTHER and BIOMASS. Prior to 2017-11-01 the BIOMASS category was understood to be a fraction of the OTHER fuel type category. The average % of the BIOMASS fraction in the sum of BIOMASS and OTHER values post 2017-11-01 was calculated as 95%. This value was therefore used to split the pre 2017-11-01 OTHER category into BIOMASS (95% of the value) and OTHER (5% of the value). In the ESPENI dataset, the 'POWER_ELEXM_OTHER_POSTCALC_MW' and 'POWER_ELEXM_BIOMASS_POSTCALC_MW' use the **calculated values** prior to 2017-11-01 and use the **reported values** for OTHER and BIOMASS **after** 2017-11-01.

A final step of the initial parsing of the files from Elexon was to create a column called ROWFLAG with all values set to 1. This was so each SDSP_RAW row had an associated value that could be manually toggled to zero during visual inspection of the data to indicate a visually identified error somewhere in that row.

**National Grid data**

These data can be manually downloaded from National Grid's historic demand data webpage[94] (https://data.nationalgrideso.com/demand/historic-demand-data) via several csv files.

The method to parse National Grid data was similar to the method for Elexon data with the creation of the SDSP_RAW column that was then used to map UTC and localtime data from the masterdatetime csv file.

The columns from the National Grid data that were used in the ESPENI dataset were:

'SETTLEMENT_DATE', 'SETTLEMENT_PERIOD', 'EMBEDDED_WIND_GENERATION', 'EMBEDDED_SOLAR_GENERATION', 'IFA_FLOW', 'IFA2_FLOW', 'BRITNED_FLOW', 'MOYLE_FLOW', 'EAST_WEST_FLOW', 'NEMO_FLOW'.

All other power columns were dropped as they would not appear in the merged ESPENI dataset.

All power columns were given a prefix of 'POWER_NGEM_' and a suffix of '_MW' to denote that the column contained values for power originally from National Grid and these were in units of MW. Similar to the Elexon data parsing a final step created a column called ROWFLAG with all values set to 1, so that each SDSP_RAW row had an associated value that could be manually toggled to zero during visual inspection of the data, to indicate an error in that row.

**Error detection for Elexon and National Grid data**

The methodology for error detection for the parsed Elexon and National Grid datasets was similar. The timeseries data were visually checked by plotting on a chart and viewed a week at a time such as shown in Figure 2; the x-axis was the datetime and the y-axis was the power in MW. When an error was found in any of the fuel type values of the row the corresponding 'ROWFLAG' row was changed in the dataset from a value of 1 to a value of 0. In many cases an erroneous data point would affect several categories for the same Settlement Period and for this reason, when a row was flagged as having an error, it was subsequently used to impute new values for ALL categories in that row, rather than having an individual flag per each data point i.e., an individual flag for each row and each fuel type category. This was felt to be an acceptable trade-off between speed and accuracy.

Although this method provides a basis for error detection for visibly obvious errors such as those displayed in Figure 2 to be flagged it suffers from three significant problems:

1) Visual error detection to clean time-series is not repeatable as it is dependent on an individual's visual assessment of the error, and this may differ between individuals
2) It is not transparent. In a coded deterministic form of error detection, although the values in the code may differ between individuals, at least the values used in the code can be considered
3) It is not scalable; it takes six times as long to visually clean 5-minute data in comparison to 30-minute data.

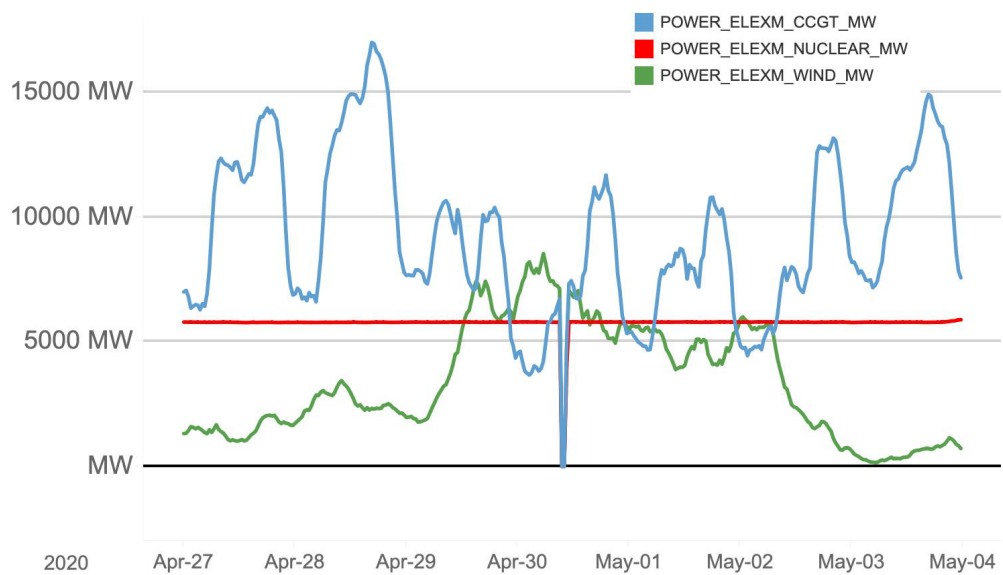

*Figure 2 – Typical error that is visually identified in the Elexon dataset where the data drops to zero for all categories for a few Settlement Periods on the 30th of April 2020*

**Error imputation**

Once each dataset had been visually checked and the erroneous rows indicated by toggling the value in the ROWFLAG column from 1 to 0, the ROWFLAG values were used to delete the MW power values across the entire row; this therefore left gaps in the dataframe across the rows manually flagged as having an error. A further step then imputed values back into the missing rows by linearly interpolating between the nearest earlier and later non-zero ROWFLAG values in that fuel type column. This method was able to accommodate ROWFLAG errors that occurred in blocks, with the longest error block having a length of 9. This method would be far less appropriate if blocks of missing data were of much greater length.

**Merging datasets and calculation of ESPENI value**

The 'SDSP_RAW' text-based column in both the Elexon and National Grid cleaned datasets was then used to merge the two datasets together into a single dataframe. In order to provide an approximation of Great Britain's electrical demand rather than simply its generation, the net imports are added to the total generation. The resulting half-hourly timeseries column termed 'ESPENI' is an acronym of **E**lexon **S**um **P**lus **E**mbedded **N**et **I**mports and is felt to represent the closest approximation to the electrical demand of Great Britain from the available public data.

# Data Records

**Raw ESPENI dataset**

An EPSENI dataset with ISO 8601 compatible datetimes for UTC and localtime and with the error flags toggled, but with the original erroneous data left in place can be downloaded at https://www.zenodo.org/record/4739408[2]. This has a filename of **espeni_raw.csv**, is Version 5, and has an md5 message-digest algorithm value of 'eb45d889ccc8031dff8e1e890ddfe70f'. It is a comma separated file in Utf-8 format and as this is the raw data pre-error correction it can be used for training of other methods to identify errors.

**Cleaned ESPENI dataset**

An EPSENI dataset with ISO 8601 compatible datetimes for UTC and localtime and with the error flags toggled, and with the original erroneous data deleted and replaced by imputed data can be downloaded at https://www.zenodo.org/record/4739408[2]. This has a filename of **espeni.csv**, is Version 5, and has an md5 message-digest algorithm value of 'ccf87847cca769a32bc60198fffc3b51'. It is a comma separated file in Utf-8 format.

The 25 x column names and descriptions for both the raw and the cleaned ESPENI datasets are shown in Table 2.

*Table 2 – Column names and descriptions for calculated the espeni_raw.csv and espeni.csv files*

| Column Name | Description |
| --- | --- |
| 'ELEXM_SETTLEMENT_DATE' | Date column format: 2020-03-30 |
| 'ELEXM_SETTLEMENT_PERIOD' | Values range from 01 to 50 |
| 'ELEXM_utc' | UTC datetime in ISO 8601 format: 2010-01-01T00:00:00+00:00 |
| 'ELEXM_localtime' | Localtime datetime in ISO 8601 format (subject to British summer time): 2010-06-01T00:00:00+01:00 |
| 'ELEXM_ROWFLAG' | Error flag row: 1 or 0; 0 indicates an error |
| 'NGEM_ROWFLAG' | Error flag row: 1 or 0; 0 indicates an error |
| 'POWER_ESPENI_MW' | Sum of all other columns prefixed with POWER, integer value in MW. Always positive. |
| 'POWER_ELEXM_CCGT_MW' | Elexon fuel type power, integer value in MW, always positive |
| 'POWER_ELEXM_OIL_MW' | Elexon fuel type power, integer value in MW, always positive |
| 'POWER_ELEXM_COAL_MW' | Elexon fuel type power, integer value in MW, always positive |
| 'POWER_ELEXM_NUCLEAR_MW' | Elexon fuel type power, integer value in MW, always positive |
| 'POWER_ELEXM_WIND_MW' | Elexon fuel type power, integer value in MW, always positive |
| 'POWER_ELEXM_PS_MW' | Elexon fuel type power, integer value in MW, always positive |
| 'POWER_ELEXM_NPSHYD_MW' | Elexon fuel type power, integer value in MW, always positive |
| 'POWER_ELEXM_OCGT_MW' | Elexon fuel type power, integer value in MW, always positive |
| 'POWER_ELEXM_OTHER_POSTCALC_MW' | Elexon fuel type power, integer value in MW, always positive |
| 'POWER_ELEXM_BIOMASS_POSTCALC_MW' | Elexon fuel type power, integer value in MW, always positive |
| 'POWER_NGEM_EMBEDDED_SOLAR_GENERATION_MW' | National Grid fuel type power, integer value in MW, always positive |
| 'POWER_NGEM_EMBEDDED_WIND_GENERATION_MW' | National Grid fuel type power, integer value in MW, always positive |
| 'POWER_NGEM_BRITNED_FLOW_MW' | National Grid interconnector power, integer value in MW, positive or negative value |
| 'POWER_NGEM_EAST_WEST_FLOW_MW' | National Grid interconnector power, integer value in MW, positive or negative value |
| 'POWER_NGEM_MOYLE_FLOW_MW' | National Grid interconnector power, integer value in MW, positive or negative value |
| 'POWER_NGEM_NEMO_FLOW_MW' | National Grid interconnector power, integer value in MW, positive or negative value |
| 'POWER_NGEM_IFA_FLOW_MW' | National Grid interconnector power, integer value in MW, positive or negative value |
| 'POWER_NGEM_IFA2_FLOW_MW' | National Grid interconnector power, integer value in MW, positive or negative value |

**Masterdatetime dataset**

The masterdatetime Python code and datetime data are available from github via the link: https://github.com/iagw/masterdatetime.git where a checksum file can be used to validate both the code and the data. The data can also be download from Zenodo[98] as a masterdatetime_iso8601.csv file in Utf-8 format, and the column headers and descriptions are shown in Table 3.

The code generates data that are then used to map the SDSP_RAW columns in the Elexon and National Grid datasets to create UTC and a localtime data. This can be used for any dataset that has Great Britain's Settlement Date and Settlement Period data, that requires an ISO 8601 datetime format. There are four Boolean columns of data that flag as true if the datetime value happens to be true, e.g., if the datetime value happens to be in a short day (any of the times in the short day) then the Boolean will show True.

*Table 3 - Column names and initial values in the masterdatetime_iso8601.csv dataset*

| Column Name | Description |
|---:|---:|
| datesp | Text value combining the settlementdate and settlementperiod values with an underscore 2000-01-01_01 |
| settlementdate | Text value of date in yyyy-mm-dd format: 2000-01-01 |
| settlementperiod | Text value of 2-digit settlement period: 01 |
| utc | UTC datetime in ISO 8601 format: 2001-01-01T00:00:00+00:00 |
| localtime | Localtime datetime in ISO 8601 format: 2001-06-01T00:00:00+01:00 |
| localtimeisdst | Boolean flag to denote whether localtime is in daylight savings: FALSE |
| short_day_flag | Boolean flag to denote whether the day is a short day: FALSE |
| long_day_flag | Boolean flag to denote whether the day is a long day: FALSE |
| normal_day_flag | Boolean flag to denote whether the day is a normal day: TRUE |

## Copyright, Disclaimer and Reservation of Rights

Although the Elexon data are publicly available (after registering through Elexon's website), it is important to note there are copyright and disclaimers associated with Elexon's data. This can be read on the Elexon webpage 'Copyright, Disclaimer and Reservation of Rights'[99] part of which states:

*No Warranty*

*No representation, warranty or guarantee is made that the information accessible via this website, or any website with which it is linked, is accurate, complete or current. No warranty or representation is made as to the availability of this site or that the functions used or materials accessible or downloaded from this site will be uninterrupted or free of errors, viruses or other harmful components. To the fullest extent permitted by law, in no event shall Elexon Limited be liable for any errors, omissions, misstatements or mistakes in any information contained on this site or damages resulting from the use of this site or any decision made or action taken in reliance of information contained in this site.*

*Use*

*The materials contained on this site are available to browse and download for the purpose only of the operation or participation in the electricity trading arrangements in Great Britain under the Balancing and Settlement Code. All commercial use is prohibited.*

The National Grid data also comes with licence conditions that limit its liability to errors and omissions within the data. A difference with the National Grid data however is that it is not behind a login, i.e., no pre-registration is required to download the data. National Grid's licence[100] can be viewed at https://data.nationalgrideso.com/licence part of which states the 'no warranty' clause:

*No warranty*

*The Information is licensed 'as is' and the Information Provider and/or Licensor excludes all representations, warranties, obligations and liabilities in relation to the Information to the maximum extent permitted by law.*

*The Information Provider and/or Licensor are not liable for any errors or omissions in the Information and shall not be liable for any loss, injury or damage of any kind caused by its use. The Information Provider does not guarantee the continued supply of the Information.*

The ESPENI data presented through this methodology from the Energy Informatics Group at the University of Birmingham continues this 'No warranty' clause from Elexon and National Grid by publishing the ESPENI under a creative commons licence CC-BY-NC 4.0[101]. This licence was chosen as it is compatible with the no warranty clauses of the Elexon and National Grid data and the non-commercial clause of the Elexon data. Choosing a creative commons licence for data encourages greater re-use of data and should be a default for all national level publicly available datasets. We would also strongly argue for sunset clauses to be mandated by regulators in all countries to make national level data CC-BY-4.0 or CC-0 after a certain period of time, e.g., after a year. This is due to the importance of historical data for modelling communities in particular, who require access to generation, energy flow, energy demand and price data that is unencumbered by licence clauses that restrict or confuse their further use. Data providers should therefore be forced to make data more available through the use of standard open data licences, by providing the data for zero or low-cost, and providing the data in a more useable format, i.e., through APIs and with ISO 8601 compatible datetimes. A sunset clause would seem to be worth exploring through regulation, so that data providers still retain the ability to leverage value from their data over a period of time, but once that time-period has lapsed, that the data becomes a public good.

ESPENI datasets created using the method presented in this article.

**Data Errors**

Table 4 indicates there was no discernible trend to the occurrence of the visually detected errors across the years of analysis, and that the errors were a small percentage of the Elexon data, and with no visual errors detected within the National Grid data. This again points to the high quality of the raw data from both Elexon and National Grid.

*Table 4 - Annual count of errors in the Elexon and National Grid raw data till 2021-05-04*

| Year | Errors flagged in Elexon data | Errors flagged in National Grid data |
|---|---|---|
| 2008 | 18 | 0 |
| 2009 | 54 | 0 |
| 2010 | 45 | 0 |
| 2011 | 27 | 0 |
| 2012 | 26 | 0 |
| 2013 | 11 | 0 |
| 2014 | 27 | 0 |
| 2015 | 14 | 0 |
| 2016 | 25 | 0 |
| 2017 | 12 | 0 |
| 2018 | 22 | 0 |
| 2019 | 29 | 0 |
| 2020 | 25 | 0 |
| Total | 335 | 0 |
| Number of rows | 219028 | 219028 |
| % Errors flagged in raw data | 0.15% | 0% |

As of 2021-05-04, the 335 flagged errors in the Elexon raw data were contained within 155 separate blocks of errors. The count of the occurrences are detailed against the length of these blocks (the sequential number of errors) in Table 5, i.e., there were 76 occurrences of a single error, and only one occurrence of 9 errors sequentially in neighbouring time periods.

Table 5 - Count of occurrences versus the length of the sequential block of errors for Elexon raw data till 2021-05-04

| Length of error block in raw Elexon data | Count of occurrences |
|---|---|
| 1 | 76 |
| 2 | 32 |
| 3 | 17 |
| 4 | 15 |
| 5 | 11 |
| 6 | 1 |
| 7 | 2 |
| 8 | 0 |
| 9 | 1 |
| Total | 147 blocks |

## Code Availability

The code for creating the masterlocaltime_iso8601.csv is available through github[102] via the link: https://github.com/iagw/masterdatetime. This also contains a yaml environment file with the details of the Python version (3.8) and the other Python packages needed to run the masterdatetime code. There is also a checksum file with shasum -256 hash values to compare against the correct versions of the data file and code.

The parsing of Elexon and National Grid to an ESPENI raw dataset (with no toggled flag values or interpolation of toggled flag values) is available through github[103] https://github.com/iagw/espeni_raw. There is also a checksum file with a shasum -256 hash value to compare against the correct version of the code.

Any codebase queries can be directed to Dr Grant Wilson: i.a.g.wilson@bham.ac.uk

## Discussion

The ESPENI values for electrical demand were compared with monthly aggregate data from the Department of Business, Energy and Industrial Strategy (BEIS). The Energy Trends *Availability and consumption of electricity (ET 5.5 - monthly)*[104] dataset was used for comparison as it is available at a monthly level. We consider the BEIS values to be the most accurate national electrical demand data at a monthly level as it contains metered data that is NOT part of the public distribution system such as from auto generators and combined heat and power plants. The BEIS data contains columns for: the amount of electricity available to the public distribution system; the transmission distribution and other losses; and then the sales of electricity in England and Wales, Scotland, and Northern Ireland. Northern Ireland data is not part of ESPENI data, as it is not part of Great Britain's electrical balancing market. Northern Ireland is part of the Integrated Single Electricity Market[105].

Although the BEIS data are more complete, it is not of sufficient granularity to analyse the changing levels of generation at a sub-monthly level; researchers have therefore typically turned to the publicly available Elexon data. Although this has a Settlement Period (30-minute) granularity, it suffers from only reporting the transmission connected generation.

Figure 3 (a) shows the growing disparity between the monthly BEIS data (the black line) and the monthly aggregated publicly available Elexon data (the blue line) since 2013. As expected, the ESPENI data (the orange line) that is the focus of the method presented in the article typically tracks the BEIS data between the Elexon data and the BEIS data due to the inclusion of embedded solar and wind generation from the publicly available National Grid data. Figure 3 (b) shows the absolute difference in Figure 3 (a) in % values against the monthly BEIS values.

*Table 6 - Annual average ESPENI and Elexon values as a % difference from BEIS annual values*

|  | 2009 | 2010 | 2011 | 2012 | 2013 | 2014 | 2015 | 2016 | 2017 | 2018 | 2019 | 2020 |
|---|---|---|---|---|---|---|---|---|---|---|---|---|
| ESPENI DIFF | -1.5% | -1.2% | -1.9% | -1.8% | -1% | -0.5% | -1.3% | -1.8% | -0.8% | -0.5% | -1.8% | -3.6% |
| Elexon DIFF | -1.6% | -1.3% | -2.4% | -2.5% | -2.5% | -3.3% | -7.2% | -7.8% | -7.8% | -8.5% | -9.8% | -12.2% |

Table 6 details these in annual terms, i.e., in 2020 the ESPENI values were on average -3.6% lower than the BEIS values, whereas the Elexon values were -12.2% lower.

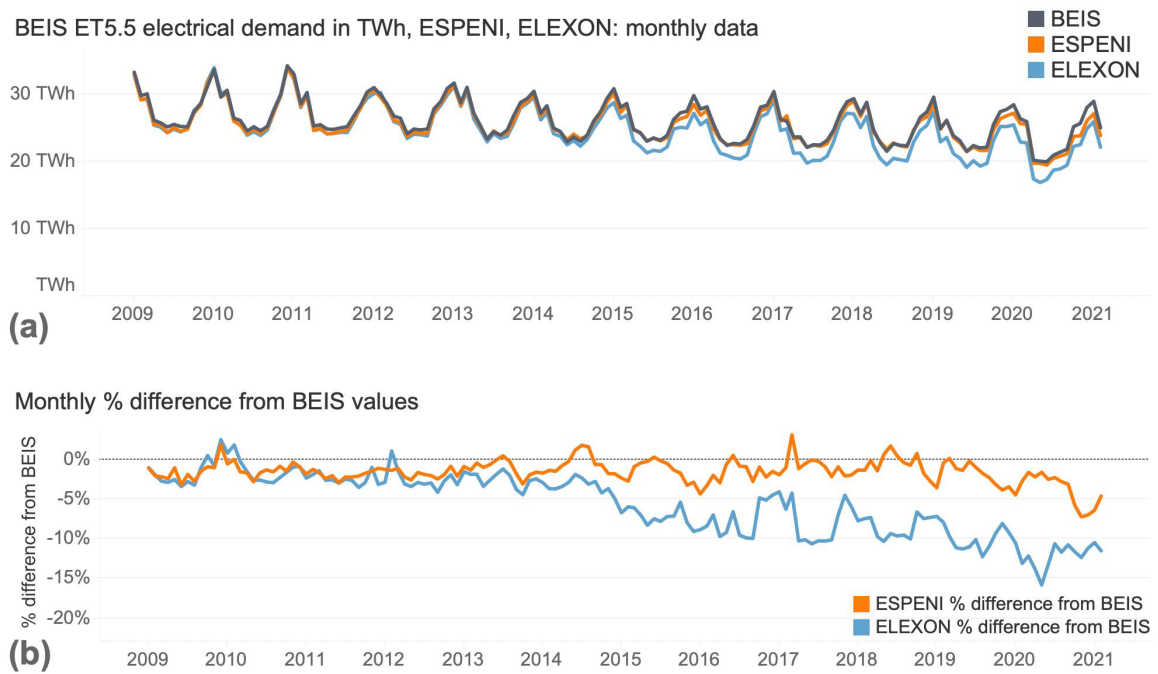

*Figure 3 – Monthly value (a) and percentage comparison (b) for BEIS (ET5.5 electrical demand), ESPENI and Elexon aggregated generation datasets*

As embedded generation continues to increase as part of the strategy to decarbonise Britain's electrical mix, the disparity between BEIS data and Elexon aggregated data will continue to grow. Figure 3 (b) and Table 6 show that ESPENI data have consistently tracked the BEIS data more closely, with the annual levels staying within 4% of the BEIS values. ESPENI data is therefore a closer approximation of the BEIS data than the Elexon data alone, which evidences the clear benefit of the method presented here.

Interestingly, from October 2020 to February 2021 inclusive, the ESPENI data has deviated from BEIS values by an average of 6.2% which is greater than any previous 5-month spell. This could just be a typical error due to the most recent months of BEIS data that becomes less pronounced as the previously estimated demand data for domestic properties are replaced by metered values. However, if this was the case, one might expect the Elexon data to also deviate from the BEIS values, whereas the Elexon data seems rather flat over the period. Therefore, at face value, something has changed over the October 2020 to February 2021 period that might suggest an underreporting of embedded wind or solar data (although the data seems on trend with previous years), or an overreporting of electrical demand from BEIS. The deviation could also be related to the different patterns of electrical demand due to Covid-19 measures. Although further detailed analysis to understand the cause of the growth in deviation over winter 2021 is not the focus of this article, it does further highlight the importance of having ESPENI in addition to Elexon and BEIS data to provide evidence for this type of detailed analysis.

**Changes to generation fuel types**

An additional benefit of the ESPENI dataset is that it allows analysis of the changing fuel types down to a half-hourly granularity. Figure 1 shows the changing fuel types at an annual level and we can clearly see the difference in the Combined Cycle Gas Turbines Natural Gas fraction that benefitted from a large change from coal between 2015 and 2016 [84]. Additional analysis could seek to combine monthly gas import data from BEIS [106] with the generation from gas to consider how the import dependency of gas generation has changed since 2009. Elexon data itself would allow this type of analysis but would underestimate the total demand for electricity as it does not contain embedded wind and solar. Thus, the ESPENI dataset has a benefit when considering generation as a fraction of total demand. As detailed above, this still differs from the BEIS monthly values for electrical demand, but less so than the Elexon data.

Extending research into import dependency still further and combining this with export analysis too, the data could be used for higher-level half-hourly analysis of the changes in use of the High Voltage Direct Current (HVDC) interconnectors that connect Great Britain to France, the Netherlands, Belgium, Northern Ireland and the Republic of Ireland. Interconnectors are viewed as an increasingly important element of Great Britain's wider energy system (not just electrical system) to integrate greater amounts of low-carbon generation and accelerate Britain's net-zero transition[107–113].

## Conclusion

Here we present a transparent repeatable methodology to create a half-hourly dataset for Great Britain's electrical demand available through its public distribution system: a half-hourly timeseries we term ESPENI. This enables analysis of Great Britain's public distribution electrical demand and the supply of this from different fuel types and interconnectors. The ESPENI timeseries tracks the official monthly values from BEIS typically within 4% of the BEIS values and is a significant improvement in granularity from monthly values to half-hourly values. The methodology could be extended down to 5-minute granularity for even more detailed analysis using Elexon 5-minute values and up sampling of the National Grid embedded solar and wind data.

This type of methodology that includes both transmission connected, and distribution connected generation to give a more complete estimate of actual electrical demand is important for other countries. Although the method presented here focusses on Great Britain's electrical market and its data, it is the principle of using all available data to infer demand by summing transmission and distribution generation and then netting off exports from imports that provides a transparent and repeatable method for other markets. Having sub-daily levels of data granularity are important for any electrical market as they attempt to decarbonise. Having this data publicly available and ideally under an open licence such as CC-BY-4.0 allows greater international comparisons of this important energy vector where the speed of decarbonisation will be an important consideration, as well as the management of flexibility over different timescales.

The major drawback of the methodology presented is at the error detection stage, which is neither fully repeatable nor scalable. The raw data has therefore been published with the errors having been flagged, but the original erroneous values being left prior to imputation to remove the errors. These two datasets may allow other researchers in different disciplines to automate the error detection and the error imputation. As more granular data becomes more available from Distribution Network Operators, Gas Distribution Networks and smart meter data, it is necessary to develop robust and repeatable methods of error detection and imputation.

Over 12 years' worth of data are published alongside the methodology[2], which provides a valuable resource for other researchers interested in wider energy strategies for Great Britain or to compare to other countries, or to allow particular types of errors to be considered for detection and imputation. The dataset shows how Great Britain's annual electrical demand has consistently fallen between 2010 and 2020, and the shift in the generation mix over this timeframe. This type of operational Settlement Period data are helpful to monitor and inform energy strategies; system operators in all countries should therefore be encouraged to make this level of quality and granularity of generation data publicly available.

## Data Availability

The ESPENI data is available under a Creative Commons Attribution Non-Commercial 4.0 International licence from https://doi.org/10.5281/zenodo.3884858. This defaults to the most recent version of the data.

## Acknowledgements

Funding: This work has been developed over several years with the support of a number of different projects including the UK Energy Research Centre research programme (grant number: EP/L024756/1) and the UK EPSRC The Active Building Centre (grant number: EP/S016627/1).

# Appendix – espeni_raw.py code

https://github.com/iagw/espeni_raw

```python
# readme
# code to create parse ESPENI dataset from publicly available Elexon and National Grid data
# to be used once Elexon and National Grid files have been manually downloaded from:
# www.elexonportal.co.uk/fuelhh (needs Elexon registration to login)
# https://data.nationalgrideso.com/demand/historic-demand-data
# elexon_data is saved to elexon_manual folder path on local machine
# national grid data is saved to national grid folder path on local machine
# masterlocaltime.csv from https://zenodo.org/record/3887182 needs to be downloaded and saved
# in 'out' folder

import os.path
import numpy as np
import datetime as dt
import pandas as pd
import glob
import time

from pandas import DataFrame

start_time = time.time()
downloaddate = dt.datetime.now().strftime("%Y-%m-%d")
yearstr = time.strftime('%Y')
open_toggle = 1

# folders
home_folder = os.getenv('HOME')
working_folder = f'{home_folder}/OneDrive - University of Birmingham/elexon_ng_espeni/'  # change to own working folder
elexon = f'{working_folder}elexon/elexon_download_data/'  # location of raw Elexon files
ngembed = f'{working_folder}ngembed/'
ngembedrawraw = f'{ngembed}ngembedrawraw/'  # location of raw national grid files
```

```python
ngembedrawpar = f'{ngembed}ngembedrawpar/'  # location of parsed
national grid files
out = f'{working_folder}/'  # location of output files

# section to parse elexon data together and add utc and localtime
#  this chooses files in a folder where the filenames contain a
particular string
#  append all of these together
os.chdir(elexon)
dfelexon = pd.DataFrame([])
for counter, file in enumerate(glob.glob('*.csv')):
    namedf = pd.read_csv(file, skiprows=0, header=0, encoding='Utf-8')
    dfelexon = dfelexon.append(namedf, sort=True)

# uppercase for all columns and strip whitespace
dfelexon = dfelexon.rename(columns=lambda x:
x.upper().strip().replace('#', '').replace(' ', '_'))
dfelexon['SETTLEMENT_PERIOD'] =
dfelexon['SETTLEMENT_PERIOD'].astype(str).str.zfill(2)
# SDSP = SETTLEMENT_DATE, SETTLEMENT_PERIOD: used as a check for
duplicates and merging
dfelexon['SDSP_RAW'] = dfelexon['SETTLEMENT_DATE'] + '_' +
dfelexon['SETTLEMENT_PERIOD']
dfelexon.duplicated(subset=['SDSP_RAW'], keep='last').sum()
dfelexon.sort_values(by=['SDSP_RAW'], ascending=True, inplace=True)
dfelexon.reset_index(drop=True, inplace=True)

# load masterlocaltime.csv file with datetimes against date and
settlement period
os.chdir(out)
masterlocaltime = pd.read_csv('masterlocaltime_iso8601.csv',
encoding='Utf-8', dtype={'settlementperiod': str})
localtimedict = dict(zip(masterlocaltime['datesp'],
masterlocaltime['localtime']))
localtimedictutc = dict(zip(masterlocaltime['datesp'],
masterlocaltime['utc']))
dfelexon['localtime'] = dfelexon['SDSP_RAW'].map(localtimedict)
dfelexon['utc'] = dfelexon['SDSP_RAW'].map(localtimedictutc)
dfelexon['SETTLEMENT_DATE'] = dfelexon['SDSP_RAW'].map(lambda x:
x.split('_')[0])
dfelexon['SETTLEMENT_PERIOD'] = dfelexon['SDSP_RAW'].map(lambda x:
x.split('_')[1])
```

```python
dfelexon['ROWFLAG'] = '1'

dfelexonlist = ['SETTLEMENT_DATE',
                'SETTLEMENT_PERIOD',
                'SDSP_RAW',
                'ROWFLAG',
                'localtime',
                'utc',
                'CCGT',
                'OIL',
                'COAL',
                'NUCLEAR',
                'WIND',
                'PS',
                'NPSHYD',
                'OCGT',
                'OTHER',
                'BIOMASS',
                'INTELEC',
                'INTEW',
                'INTFR',
                'INTIFA2',
                'INTIRL',
                'INTNED',
                'INTNEM',
                'INTNSL']

dfelexon['utc'] = pd.to_datetime(dfelexon['utc'])
dfelexon = dfelexon.set_index('utc', drop=False)

dfelexon = dfelexon[dfelexonlist]

# renames column names
nosuffix = ['SETTLEMENT_DATE', 'SETTLEMENT_PERIOD', 'localtime', 'utc', 'ROWFLAG', 'SDSP_RAW']
suffix = 'ELEXM'
for col in dfelexon.columns:
    if col in nosuffix:
        dfelexon.rename(columns={col: f'{suffix}_{col}'}, inplace=True)
```

```python
    else:
        dfelexon.rename(columns={col: f'POWER_{suffix}_{col}_MW'}, inplace=True)

# section to parse national grid data together and add utc and localtime
os.chdir(ngembedrawraw)
existing_downloaded_rawraw_files = pd.DataFrame(glob.glob('*.csv'))
os.chdir(ngembedrawpar)
existing_parsed_raw_files = pd.DataFrame(glob.glob('*.csv'))
if existing_parsed_raw_files.empty:
    csvs_to_parse = existing_downloaded_rawraw_files[0]
else:
    existing_parsed_raw_files[0] = existing_parsed_raw_files[0].map(lambda x: x.split('_rawpar.csv')[0] + '_rawraw.csv')
    csvs_to_parse = existing_downloaded_rawraw_files[0][~existing_downloaded_rawraw_files[0].isin

(existing_parsed_raw_files[0])]

for fname in csvs_to_parse:
    os.chdir(ngembedrawraw)
    # then reads in the demandupdate file (the one that changes on a daily basis)
    df = pd.read_csv(fname, encoding='Utf-8')
    if 'FORECAST_ACTUAL_INDICATOR' in df.columns:
        df = df[df['FORECAST_ACTUAL_INDICATOR'] != 'F']
        df = df.drop('FORECAST_ACTUAL_INDICATOR', axis=1)

    # uppercase for all columns and strip whitespace
    df = df.rename(columns=lambda x: x.upper().strip())

    # get date from SETTLEMENT_DATE column
    # date format starts with lowercase month in format 01-Apr-2005
    # but changes to uppercase month text in recent years 01-APR-2015
    # therefore change everything to uppercase and change to datetime to standardise
    df['SETTLEMENT_DATE'] = pd.to_datetime(df['SETTLEMENT_DATE'].squeeze().str.upper().tolist(), format='%d-%b-%Y')
```

```python
    # SETTLEMENT_DATE_TEXT and SETTLEMENT_PERIOD_TEXT created
    df['SETTLEMENT_DATE'] = df['SETTLEMENT_DATE'].dt.strftime('%Y-%m-%d')
    df['SETTLEMENT_PERIOD'] = df['SETTLEMENT_PERIOD'].astype(str).str.zfill(2)

    # SDSP = settlement date, settlement period: used as a check for duplicates etc.
    df['SDSP_RAW'] = df['SETTLEMENT_DATE'].astype(str) + '_' + df['SETTLEMENT_PERIOD'].astype(str)

    # drop all rows that have a forecast value i.e. keep all rows that do not have 'F'
    # drop FORECAST_ACTUAL_INDICATOR as column is no longer needed

    df.sort_values(['SETTLEMENT_DATE', 'SETTLEMENT_PERIOD'], ascending=[True, True], inplace=True)
    df.drop_duplicates(subset=['SDSP_RAW'], keep='first', inplace=True)
    df.reset_index(drop=True, inplace=True)

    # next line makes a list? of all columns that are NOT in the list of 'SETTLEMENT_DATE' etc.
    nosuffix = ['SETTLEMENT_DATE', 'SETTLEMENT_PERIOD', 'FORECAST_ACTUAL_INDICATOR', 'SDSP_RAW']
    for col in df.columns:
        if col in nosuffix:
            df.rename(columns={col: f'{suffix}_{col}'}, inplace=True)
        else:
            df.rename(columns={col: f'POWER_{suffix}_{col}_MW'}, inplace=True)

    os.chdir(ngembedrawpar)
    fname = fname.split('_rawraw.csv')[0] + '_rawpar.csv'
    df.to_csv(fname, encoding='Utf-8', index=False)

#  loop chooses files in a folder who's names contain a particular string
#  and appends all of these together
os.chdir(ngembedrawpar)
suffix = 'NGEM'
dfng = pd.DataFrame([])
```

```python
for counter, file in enumerate(glob.glob('*_rawpar.csv')):
    namedf = pd.read_csv(file, skiprows=0, encoding='utf-8')
    dfng = dfng.append(namedf, sort=True)

dfng[f'{suffix}_SETTLEMENT_PERIOD'] = dfng[f'{suffix}_SETTLEMENT_PERIOD'].astype(str).str.zfill(2)

dfng.sort_values([f'{suffix}_SDSP_RAW'], ascending=[True], inplace=True)
dfng.drop_duplicates(subset=[f'{suffix}_SDSP_RAW'], keep='last', inplace=True)

dfng.reset_index(drop=True, inplace=True)

# load masterlocaltime.csv file with datetimes against date and settlement period

os.chdir(out)
dfng[f'{suffix}_localtime'] = dfng[f'{suffix}_SDSP_RAW'].map(localtimedict)
dfng[f'{suffix}_utc'] = dfng[f'{suffix}_SDSP_RAW'].map(localtimedictutc)
dfng.insert(3, 'NGEM_ROWFLAG', value=1)

df = pd.merge(dfelexon, dfng, how='left', left_on=['ELEXM_SDSP_RAW'], right_on=['NGEM_SDSP_RAW'])
biomasslist = {'POWER_ELEXM_BIOMASS_MW': 'POWER_ELEXM_BIOMASS_PRECALC_MW',
               'POWER_ELEXM_OTHER_MW': 'POWER_ELEXM_OTHER_PRECALC_MW'}
df = df.rename(columns=biomasslist)

mask = df['ELEXM_utc'].astype(str).str.contains('/', regex=True)
df.loc[mask, 'ELEXM_utc'] = pd.to_datetime(df.loc[mask, 'ELEXM_utc'],
                                           format='%d/%m/%Y %H:%M').dt.strftime('%Y-%m-%d %H:%M')
df['ELEXM_utc'] = pd.to_datetime(df['ELEXM_utc'], utc=True, format='%Y-%m-%d %H:%M')

mask = df['ELEXM_SETTLEMENT_DATE'].astype(str).str.contains('/', regex=True)
df.loc[mask, 'ELEXM_SETTLEMENT_DATE'] = pd.to_datetime(df.loc[mask, 'ELEXM_SETTLEMENT_DATE'],
```

```python
                                        format='%d/%m/%Y').dt.strftime('%Y-%m-%d')
df['ELEXM_SETTLEMENT_DATE'] = pd.to_datetime(df['ELEXM_SETTLEMENT_DATE'], utc=True,
                                             format='%Y-%m-%d').dt.strftime('%Y-%m-%d')
# df['ELEXM_SETTLEMENT_DATE'] = df['ELEXM_SETTLEMENT_DATE'].dt.strftime('%Y-%m-%d')

mask = df['ELEXM_localtime'].astype(str).str.contains('/', regex=True)
df.loc[mask, 'ELEXM_localtime'] = pd.to_datetime(df.loc[mask, 'ELEXM_localtime'],
                                                  format='%d/%m/%Y %H:%M').dt.strftime('%Y-%m-%d %H:%M')
df['ELEXM_localtime'] = pd.to_datetime(df['ELEXM_localtime'], utc=True, format='%Y-%m-%d %H:%M')
df: DataFrame = df.set_index('ELEXM_utc', drop=False)

# elexonstartdate = '2008-11-06 00:00:00'
biomassstartdate = '2017-11-01 20:00:00+00:00'
otherfinishdate = '2017-11-01 20:30:00+00:00'

# sum of biomass since start date of OTHER split to OTHER and BIOMASS
bb = df['POWER_ELEXM_BIOMASS_PRECALC_MW'][biomassstartdate:].mean()
oo = df['POWER_ELEXM_OTHER_PRECALC_MW'][otherfinishdate:].mean()
otherb_ratio = oo/bb
oblist = ['POWER_ELEXM_OTHER_PRECALC_MW', 'POWER_ELEXM_BIOMASS_PRECALC_MW']
df['POWER_ELEXM_OTHER_POSTCALC_MW'] = \
    np.where(df.index < otherfinishdate, df[oblist].sum(axis=1).mul(otherb_ratio),
             df['POWER_ELEXM_OTHER_PRECALC_MW']).round(0)
df['POWER_ELEXM_BIOMASS_POSTCALC_MW'] = \
    np.where(df.index < biomassstartdate, df[oblist].sum(axis=1).mul(1-otherb_ratio),
             df['POWER_ELEXM_BIOMASS_PRECALC_MW']).round(0)

# list to sum to ESPENI
espenilist = ['POWER_ELEXM_CCGT_MW',
              'POWER_ELEXM_OIL_MW',
              'POWER_ELEXM_COAL_MW',
              'POWER_ELEXM_NUCLEAR_MW',
              'POWER_ELEXM_WIND_MW',
```

```
                'POWER_ELEXM_PS_MW',
                'POWER_ELEXM_NPSHYD_MW',
                'POWER_ELEXM_OCGT_MW',
                'POWER_ELEXM_OTHER_POSTCALC_MW',
                'POWER_ELEXM_BIOMASS_POSTCALC_MW',
                'POWER_NGEM_BRITNED_FLOW_MW',  # added on 2021-03-02,
INTNED_FLOW changed to BRITNED_FLOW
                'POWER_NGEM_EAST_WEST_FLOW_MW',
                'POWER_NGEM_MOYLE_FLOW_MW',
                'POWER_NGEM_NEMO_FLOW_MW',
                'POWER_NGEM_IFA_FLOW_MW',   # added on 2021-03-02,
FRENCH_FLOW changed to IFA_FLOW
                'POWER_NGEM_IFA2_FLOW_MW',   # added on 2021-03-02, IFA2
data started
                'POWER_NGEM_EMBEDDED_SOLAR_GENERATION_MW',
                'POWER_NGEM_EMBEDDED_WIND_GENERATION_MW']

df['POWER_ESPENI_MW'] = df[espenilist].sum(axis=1)

espenifileoutput = ['ELEXM_SETTLEMENT_DATE',
                    'ELEXM_SETTLEMENT_PERIOD',
                    'ELEXM_utc',
                    'ELEXM_localtime',
                    'ELEXM_ROWFLAG',
                    'NGEM_ROWFLAG',
                    'POWER_ESPENI_MW',
                    'POWER_ELEXM_CCGT_MW',
                    'POWER_ELEXM_OIL_MW',
                    'POWER_ELEXM_COAL_MW',
                    'POWER_ELEXM_NUCLEAR_MW',
                    'POWER_ELEXM_WIND_MW',
                    'POWER_ELEXM_PS_MW',
                    'POWER_ELEXM_NPSHYD_MW',
                    'POWER_ELEXM_OCGT_MW',
                    'POWER_ELEXM_OTHER_POSTCALC_MW',
                    'POWER_ELEXM_BIOMASS_POSTCALC_MW',
                    'POWER_NGEM_EMBEDDED_SOLAR_GENERATION_MW',
                    'POWER_NGEM_EMBEDDED_WIND_GENERATION_MW',
                    'POWER_NGEM_BRITNED_FLOW_MW',  # added on 2021-
03-02, INTNED_FLOW changed to BRITNED_FLOW
                    'POWER_NGEM_EAST_WEST_FLOW_MW',
```

```
                        'POWER_NGEM_MOYLE_FLOW_MW',
                        'POWER_NGEM_NEMO_FLOW_MW',
                        'POWER_NGEM_IFA_FLOW_MW',  # added on 2021-03-02, FRENCH_FLOW changed to IFA_FLOW
                        'POWER_NGEM_IFA2_FLOW_MW',  # added on 2021-03-02, IFA2 data started
                        ]

df = df[espenifileoutput]
os.chdir(out)
# df.to_csv('espeni_raw.csv', encoding='Utf-8', index=False)
print("time elapsed: {:.2f}s".format(time.time() - start_time))
```